# Predicting membrane protein contacts from non-membrane proteins by deep transfer learning

**Zhen Li[1,3,*], Sheng Wang[1,2,*], Yizhou Yu[3], Jinbo Xu[1,†]**

[1]Toyota Technological Institute at Chicago

[2]Department of Human Genetics, University of Chicago

[3]Department of Computer Science, University of Hong Kong

* These authors contributed equally to the work as first authors.

† To whom correspondence should be addressed.

## Abstract

Computational prediction of membrane protein (MP) structures is very challenging partially due to lack of sufficient solved structures for homology modeling or parameter estimation of computational methods. Recently direct evolutionary coupling analysis (DCA) sheds some light on protein contact prediction and accordingly, contact-assisted folding, but DCA is effective only on some very large-sized families since it uses information only in a single protein family. This paper presents a deep transfer learning method that can significantly improve MP contact prediction by learning contact occurring patterns and complex sequence-contact relationship from thousands of non-membrane proteins (non-MPs). Tested on 510 non-redundant MPs, our deep model (learned from only non-MPs) has top L/10 long-range contact prediction accuracy 0.69, better than our deep model trained by only MPs (0.63) and much better than a representative DCA method CCMpred (0.47) and the CASP11 winner MetaPSICOV (0.55). The accuracy of our deep model can be further improved to 0.72 when trained by a mix of non-MPs and MPs. When only contacts in transmembrane regions are evaluated, our method has top L/10 long-range accuracy 0.62, 0.57, and 0.53 when trained by a mix of non-MPs and MPs, by non-MPs only, and by MPs only, respectively, still much better than MetaPSICOV (0.45) and CCMpred (0.40). All these results suggest that sequence-structure relationship learned by our deep model from non-MPs generalizes well to MP contact prediction. Improved contact prediction also leads to better contact-assisted folding. Using only top predicted contacts as restraints (i.e., no force fields used), our deep learning method can fold 160 and 200 of 510 MPs with TMscore>0.6 when trained by non-MPs only and by a mix of non-MPs and MPs, respectively, while CCMpred and MetaPSICOV can do so for only 56 and 77 MPs, respectively. Our contact-assisted folding also greatly outperforms homology modeling, which can fold 41 MPs with TMscore>0.6 when a mix of MPs and non-MPs are used as templates and 3 MPs when only non-MPs are used as templates.

## Introduction

It is well known that membrane proteins (MPs) are important for drug design and have been targeted by approximately half of current therapeutic drugs [6]. MPs are also important in cell-environment interactions by serving as environment sensing receptors, transporters, and channels. It is estimated that 20-40% of all genes in most genomes encode MPs [3, 4] and larger genomes contain a higher fraction of MPs [5]. In particular, the Human genome has >5,000 reviewed MPs and more than 3000 of them are non-redundant (when 25% sequence identity is used as cutoff to remove redundancy). To better understand MPs, it is better to have their 3D structures. Experimental determination of MP structures is challenging as they are often too large for nuclear magnetic resonance (NMR) experiments and very difficult to crystallize [1]. Currently there are only about 510 non-redundant MPs with solved structures [2]. That is, most MPs have no solved structures and it is very important to develop computational methods to predict MP structures from sequence information.

Although knowledge-based computational structure prediction works well for a large number of globular proteins, it faces some challenges for MPs partially due to lack of MPs with solved structures. First, since there are not many solved MP structures as templates, many MPs cannot be modelled by homology modeling. As such, we have to resort to de novo prediction or ab initio folding. Second, accurate estimation of the parameters of a computational method for MP structure prediction rely on sufficient statistics, which sometimes cannot be fulfilled due to a small number of non-redundant MPs with solved structures. So far the most successful methods for ab initio folding are based upon fragment assembly, but it works on only some small proteins. Very recently, contact-assisted ab initio folding has made good progress for some proteins with a large number (>1000) sequence homologs. This technique first predicts the contacts of a protein in question and then use predicted contacts as restraints to guide ab initio folding. However, contact-assisted folding heavily depends on accurate prediction of protein contacts.

There are a few slightly different definitions for contacts. Here we define that two residues form a contact if in native, the Euclidean distance of their $C_\beta$ atoms is less than 8Å. Evolutionary coupling analysis (ECA) and supervised learning are two popular contact prediction methods. ECA predicts contacts by identifying co-evolved residues, such as EVfold [7], PSICOV [8], CCMpred [9] and Gremlin [10], but it needs a large number of sequence homologs to be effective [11, 12]. Supervised learning predicts contacts from a variety of information, e.g., SVMSEQ[13], CMAPpro [14], PconsC2 [11] and PconsC3 [15], MetaPSICOV [16], PhyCMAP [17] and CoinDCA-NN [12]. CCMpred is one of the best ECA methods and MetaPSICOV is the CASP11 winners among all methods. CMAPpro [14] uses a deep architecture, but its performance saturates at ~10 layers and was worse than MetaPSICOV (a 2-layer neural network) in CASP11 [33]. There are some contact prediction methods which are specifically developed for MPs. These methods employ some MP-specific features and train their parameters mainly from MPs, so they may perform slightly better than the abovementioned ECA or supervised methods. For example, TMHcon [20] is a neural network method that integrates MP-specific and -independent features to predict MP inter-helical contacts. McAllister and Floudas [21] proposed a mixed integer programming method for MP contact prediction. TMhit [22] combines a two-level hierarchical scheme with a support vector machine (SVM) classifier. MEMPACK [23] is another SVM method that predicts contacts from sequence profile, lipid exposure, sequence separation and relative position of a residue in MP helices. Other examples include TMhhcp [24], MemBrain [25], COMSAT [26] and OMPcontact [27].

Although so many methods have been developed, for MPs with fewer sequence homologs, the predicted contacts by existing methods such as CCMpred, PSICOV, Evfold, MetaPSICOV and CoinDCA [18] are still of low quality and not very helpful to ab initio folding [19]. This motivates us to develop a better contact prediction method, especially for small-sized protein families.

In this paper we present an ultra-deep learning method for MP contact prediction and accordingly contact-assisted folding. We treat a protein contact map as an image and formulate contact prediction similarly as pixel-level image labeling. However, we cannot directly apply the models developed for image labeling to contact prediction since proteins have more complex features and the ratio of contacts (i.e., positive labels) is very small (<2%). Instead, we propose a new deep architecture for contact prediction by concatenating two deep residual neural networks [28]. Deep residual networks have won the ILSVRC and COCO 2015 competitions and now are very popular in image recognition [29, 30]. The challenge of applying deep learning to MP contact prediction is lack of sufficient training data. To overcome this, we train our deep learning model using a large number of non-MPs with solved structures. It turns out that the resultant deep model works well for MP contact prediction, outperforming our (shallower) deep model trained by only MPs and existing popular methods such as CCMpred and MetaPSICOV. Our further study indicates that using a mix of non-MPs and MPs in a good way, we can train a deep model with even better prediction accuracy, especially in transmembrane regions.

## Method

**Protein features.** Given a membrane protein (MP) sequence under prediction, we first run PSI-BLAST [31] or HHblits [32] to find its sequence homologs and then build a multiple sequence alignment (MSA) of all the sequence homologs. Starting from the MSA, we derive two types of protein features: sequential features and pairwise features, which are also called 1-dimensional (1D) and 2-dimensional (2D) features, respectively. The sequential features include protein sequence profile, predicted secondary structure (SS) [34] and solvent accessibility (SA) [35]. We predict SS and SA from sequence profile using our in-house tool RaptorX-Property [36]. The pairwise features include direct co-evolutionary information generated by CCMpred [9], mutual information and pairwise contact potential [37, 38]. Different from many existing MP contact prediction methods that use some MP-specific features, we do not use any MP-specific features since we found out that they are not very helpful to our deep learning model possibly because our model can implicitly learn these features.

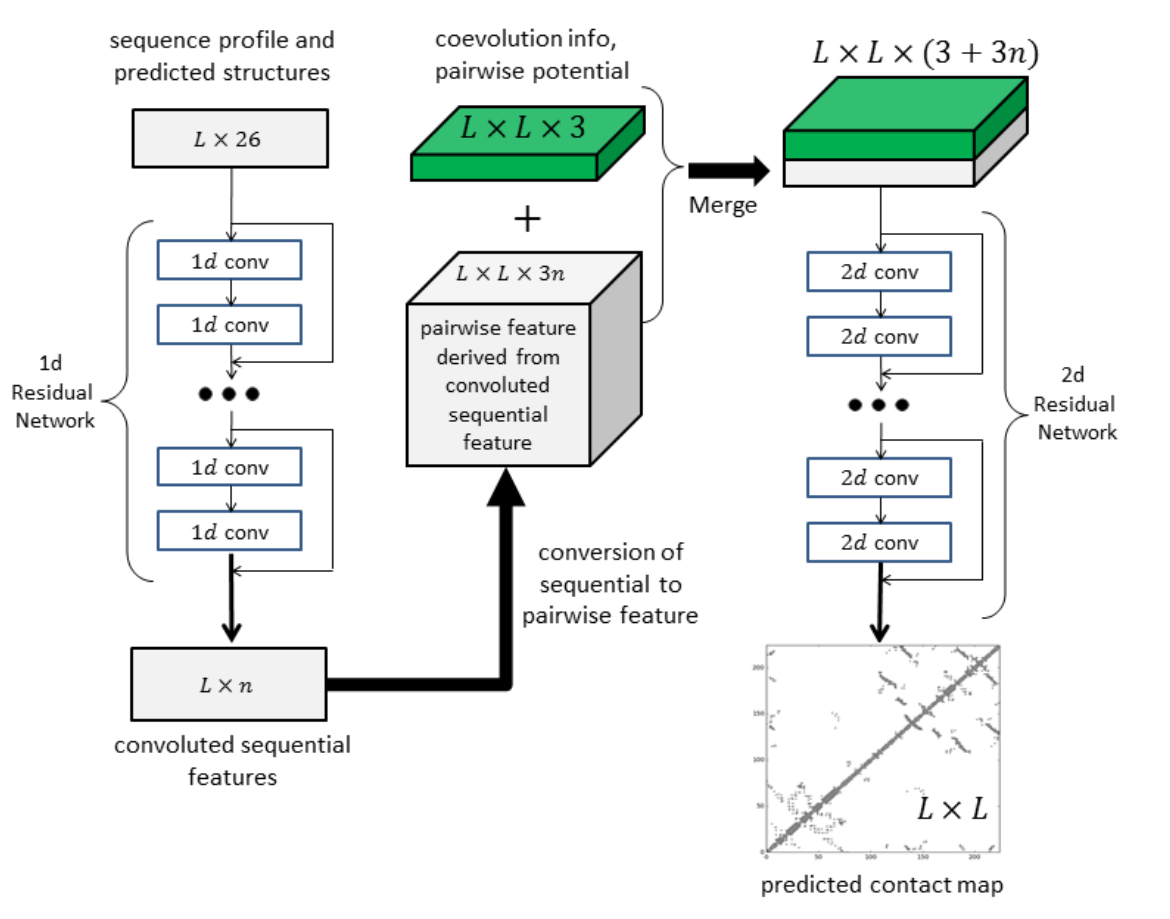

**Figure 1.** Overview of our deep learning model for MP contact prediction where L is the sequence length of one MP under prediction.

As shown in Fig. 1, our deep learning model is formed by two residual neural networks. Each residual network is composed of some residual blocks, each block in turn consisting of 2 convolution layers and 2 ReLU activation layers (Fig. 2). In the first residual network, $X_l$ and $X_{l+1}$ represent sequential features and have dimension $L \times n_l$ and $L \times n_{l+1}$, respectively, where $n_l$ ($n_{l+1}$) is the number of features (also called hidden neurons) at each residue. In the 2^nd residual network, $X_l$ and $X_{l+1}$ represent pairwise features and have dimension $L \times L \times n_l$ and $L \times L \times n_{l+1}$, respectively, where $n_l$ ($n_{l+1}$) is the number of features (hidden neurons) at one residue pair. Batch normalization is applied before each activation layer. The first residual network mainly conducts a series of 1-dimensional (1D) convolutional transformations of sequential features. Its output is converted to a 2-dimensional (2D) matrix by an operation similar to outer product and then fed into the 2nd residual network together with the pairwise features. The 2nd residual network mainly conducts a series of 2D convolutional transformations of its input. Finally, the output of the 2nd network is fed into logistic regression, which predicts the probability of any two residues in a contact. The filter size (i.e., window size) used by a 1D convolution layer is 17 while that used by a 2D convolution layer is 3×3 or 5×5. We fix the depth (i.e., the number of convolution layers) of the first residual network, but vary the depth of the second network.

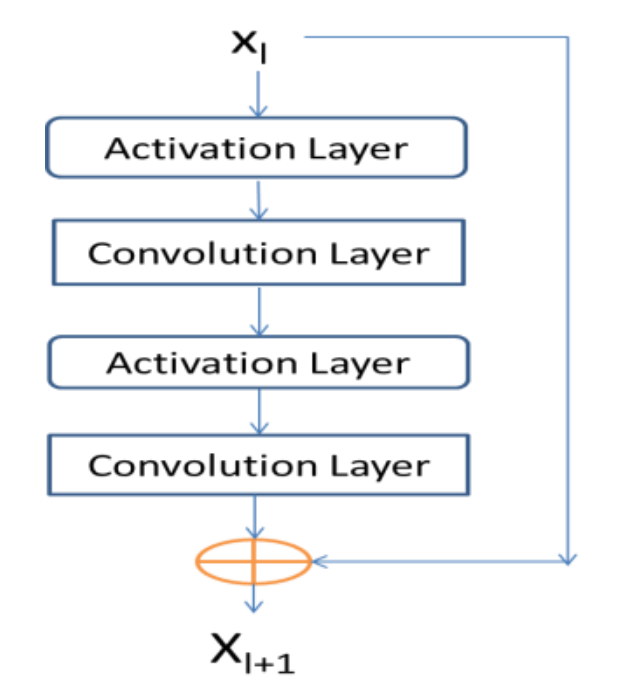

**Figure 2.** A building block of our residual network with $X_l$ and $X_{l+1}$ being input and output, respectively. Each block has 2 convolution layers and 2 activation layers.

Our deep learning method is unique in the following aspects. First, our model employs a concatenation of two deep residual neural networks, which has not been applied to contact prediction before. Second, we predict all the contacts of a protein simultaneously, as opposed to existing supervised methods that predict contacts one by one. By simultaneous prediction, we can easily learn contact occurring patterns in a protein and model complex, long-range sequence-contact correlation. A simple statistics on membrane proteins or globular proteins indicate that some contact patterns appear much more frequently than the others.

We train our models by maximum-likelihood with $L_2$-norm regularization. A stochastic gradient descent algorithm is used to minimize the objective function. The whole algorithm is implemented with Theano (http://www.deeplearning.net/software/theano/) and runs on a GPU card.

**Datasets.** In total there are 510 non-redundant MPs with solved structures from PDBTM [2]. They serve as the ground truth in evaluating the competing prediction methods. Among these 510 MPs, 5 of them have no contacts in transmembrane regions. So when evaluating prediction accuracy in transmembrane regions, only 505 MPs are used. We randomly divide these 510 MPs into 5 subsets (denoted as MP0, MP1, MP2, MP3 and MP4) of same size so that we can do 5-fold cross validation when MPs are also used in the training set. That is, we use 4 subsets in the training set and the remaining 1 subset as the test set.

So far there are more than 10,000 proteins in PDB25, a representative set of non-redundant proteins in Protein Data Bank (PDB) in which any two proteins share less than 25% sequence identity [39]. To remove redundancy between MPs and PDB25, we exclude the proteins in PDB25 sharing >25% sequence identity or having a BLAST E-value <0.1 with any of the 510 MPs. This results in 9627 non-MPs in PDB25, all of which are non-redundant to the 510 MPs. From these 9627 non-MPs, we randomly sample 600 proteins as the validation set and use the remaining ~9000 proteins as the training proteins.

**Model training strategies.** To study how to make the best use of non-MPs for MP contact prediction, we mainly consider the following three model training strategies.

1) Training deep models by MPs only (denoted as MP-only). That is, we train our deep learning model

using only the 510 MPs. In total we have trained 5 models. For each model, we use 4/5 of the 510 (i.e., 408) MPs as the training and validation proteins and the remaining 1/5 MPs as the test proteins. We form a validation set by randomly selecting 30 out of the 408 MPs and 100 non-MPs. The validation set is used to determine when to stop training and the regularization factor. Since there are only 378 MPs in the training set, we cannot use a very deep learning model. We tested quite a few architectures of our deep learning model and found out that a model with two 1D convolutional layers (and 50 hidden neurons at each layer) and twenty 2D convolutional layers (and 30 hidden neurons at each layer) yields the best performance. We use the same architecture for all the 5 models. We terminate the training procedure at 15 epochs (each epoch scans through all the training data once).

2) Training deep models by non-MPs only (denoted as NonMP-only). That is, we train our deep learning model without using any MPs. In this case, we have trained only one model using the ~9000 non-MPs and validate it using 600 non-MPs. Finally we test this model using all the 510 MPs. Since we have a large training set in this case, we use a model with six 1D convolutional layers (50 hidden neurons at each layer) and sixty 2D convolutional layers (and 60 hidden neurons at each layer). We terminate the training procedure at 20 epochs.

3) Training deep models by a mix of non-MPs and MPs (denoted as Mixed). In total we have trained 5 models. Each model is trained by the 9000 non-MPs and 4/5 of the MPs. We validate this model using the 600 non-MPs and test it using the remaining 1/5 of the MPs. Since there is a large training set, here we use the same architecture for our deep model as the NonMP-only strategy. Since there are much fewer MPs than non-MPs, to achieve the best performance, we assign different weight to non-MPs and MPs. Our experimental results (data not shown) show that by setting the weight of MPs to 5 and that of non-MPs to 1, we can obtain very good accuracy. By the way, if we set the weight of MPs to infinity, this strategy becomes the MP-only strategy. Again we terminate the training procedure at 20 epochs.

**Competing methods.** We have tested PSICOV [8], Evfold [7], CCMpred [9], and MetaPSICOV [16]. The former three methods are representative EC methods. MetaPSICOV is a supervised learning method that performed the best in CASP11 [33]. MetaPSICOV predicts contacts using as input the output of the former three programs as well as all the protein features used by our method. These four programs are run with parameters set according to their respective papers. We cannot evaluate PconsC2 [11] since we failed to obtain any results from its web server. PconsC2 did not outperform MetaPSICOV in CASP11 [33], so it may suffice to just compare our method with MetaPSICOV. We cannot evaluate PconsC3 (a slightly improved version of PconsC2) either since its web server can only predict several test proteins in a week. There are also a couple of other tools such as Dong Xu's method [27] and COMSAT [26] specifically developed for MP contact prediction. Dong Xu's method is not available for comparison due to disk failure in his group. We cannot compare our method with COMSAT since it requires us to provide extra information as input, in addition to primary sequence. Further, COMSAT predicts if there is a contact between two helix segments instead of between two residues, while our method only needs the primary sequence as input and predicts inter-residue contacts.

# Results

We say a contact is short-, medium- and long-range when the sequence distance of two residues in a contact falls into [6, 11], [12, 23], and ⩾24, respectively. We evaluate the accuracy of the top *L/k* (*k*=10, 5, 2, 1) predicted contacts where L is protein sequence length. The prediction accuracy is defined as the percentage of native contacts among the top *L/k* predicted contacts. In the case that there are no *L/k* native contacts in a category, we simply use *L/k* as the denominator when calculating the accuracy. We measure the quality of a 3D model by TMscore [41]. It ranges from 0 to 1, with 0 indicating the worst and 1 the best, respectively. A 3D model with TMscore⩾0.6 has a correct fold while a 3D model with TMscore<0.5 usually does not. Some reports also use TMscore=0.5 as a cutoff to determine if a model has a correct fold or not.

## Overall performance

Tables 1 and 2 show that the performance of our three model training strategies as well as the other competing methods. All our three model training strategies outperform the other competing methods in terms of medium- and long-range prediction accuracy. It is quite interesting that our deep model trained by only non-MPs greatly outperforms our deep model trained by only MPs even if only the predicted contacts in transmembrane regions are evaluated. This result suggests that non-MPs and MPs share some common properties for contact prediction that can be learned by our deep learning model. Our deep network obtains the best performance especially for contacts in transmembrane regions when both non-MPs and MPs are used in the training set. MetaPSICOV is also a supervised learning model mainly trained by non-MPs, but it performs much worse than our NonMP strategy because 1) it is a shallow model while our is a deep model and 2) it is trained by no more than 1000 proteins while our deep model is trained by about 9000 non-MPs.

**Table 1.** Overall contact prediction accuracy on 510 membrane proteins. 'MP', 'NonMP', and 'Mixed' indicate the models are trained by membrane proteins only, by non-membrane proteins only and by a mix of membrane and non-membrane proteins.

| **Method** | **Short** | | | | **Medium** | | | | **Long** | | | |
|---|---|---|---|---|---|---|---|---|---|---|---|---|
| | **L/10** | **L/5** | **L/2** | **L** | **L/10** | **L/5** | **L/2** | **L** | **L/10** | **L/5** | **L/2** | **L** |
| EVfold | 0.15 | 0.12 | 0.08 | 0.06 | 0.24 | 0.18 | 0.11 | 0.08 | 0.39 | 0.33 | 0.23 | 0.16 |
| PSICOV | 0.20 | 0.14 | 0.09 | 0.06 | 0.25 | 0.18 | 0.11 | 0.07 | 0.38 | 0.31 | 0.21 | 0.15 |
| CCMpred | 0.24 | 0.17 | 0.10 | 0.07 | 0.32 | 0.23 | 0.13 | 0.08 | 0.47 | 0.40 | 0.28 | 0.19 |
| MetaPSICOV | 0.40 | 0.31 | 0.19 | 0.12 | 0.43 | 0.34 | 0.23 | 0.15 | 0.55 | 0.49 | 0.37 | 0.27 |
| Our method (MP) | 0.35 | 0.27 | 0.16 | 0.10 | 0.48 | 0.37 | 0.24 | 0.16 | 0.63 | 0.57 | 0.45 | 0.33 |
| Our method (NonMP) | 0.51 | 0.39 | 0.24 | 0.14 | 0.57 | 0.45 | 0.29 | 0.18 | 0.69 | 0.65 | 0.53 | 0.40 |
| Our method (Mixed) | 0.53 | 0.40 | 0.24 | 0.14 | 0.59 | 0.47 | 0.30 | 0.19 | 0.72 | 0.68 | 0.57 | 0.43 |

**Table 2.** The prediction accuracy of contacts in transmembrane regions for the 505 membrane proteins.

| **Method** | **Short** | | | | **Medium** | | | | **Long** | | | |
|---|---|---|---|---|---|---|---|---|---|---|---|---|
| | **L/10** | **L/5** | **L/2** | **L** | **L/10** | **L/5** | **L/2** | **L** | **L/10** | **L/5** | **L/2** | **L** |
| EVfold | 0.21 | 0.16 | 0.11 | 0.08 | 0.34 | 0.27 | 0.18 | 0.12 | 0.36 | 0.31 | 0.22 | 0.16 |
| PSICOV | 0.24 | 0.19 | 0.12 | 0.09 | 0.32 | 0.24 | 0.16 | 0.11 | 0.33 | 0.27 | 0.19 | 0.14 |
| CCMpred | 0.29 | 0.22 | 0.14 | 0.09 | 0.38 | 0.30 | 0.18 | 0.12 | 0.40 | 0.34 | 0.25 | 0.17 |
| MetaPSICOV | 0.41 | 0.32 | 0.22 | 0.16 | 0.46 | 0.38 | 0.26 | 0.18 | 0.45 | 0.39 | 0.29 | 0.22 |
| Our method (MP) | 0.48 | 0.38 | 0.24 | 0.16 | 0.55 | 0.47 | 0.32 | 0.21 | 0.53 | 0.47 | 0.36 | 0.27 |
| Our method (NonMP) | 0.54 | 0.42 | 0.27 | 0.18 | 0.58 | 0.50 | 0.34 | 0.23 | 0.57 | 0.53 | 0.42 | 0.31 |
| Our method (Mixed) | 0.56 | 0.45 | 0.29 | 0.19 | 0.62 | 0.54 | 0.37 | 0.24 | 0.62 | 0.58 | 0.47 | 0.35 |

Figure 3 shows the prediction accuracy with respect to the number of non-redundant sequence homologs available for a protein under prediction. This figure indicates that our methods work particularly well when a protein in question does not have many sequence homologs. However, even if thousands of sequence homologs are available for a protein under prediction, our methods on average still perform much better than CCMpred and MetaPSICOV, although our advantage becomes smaller.

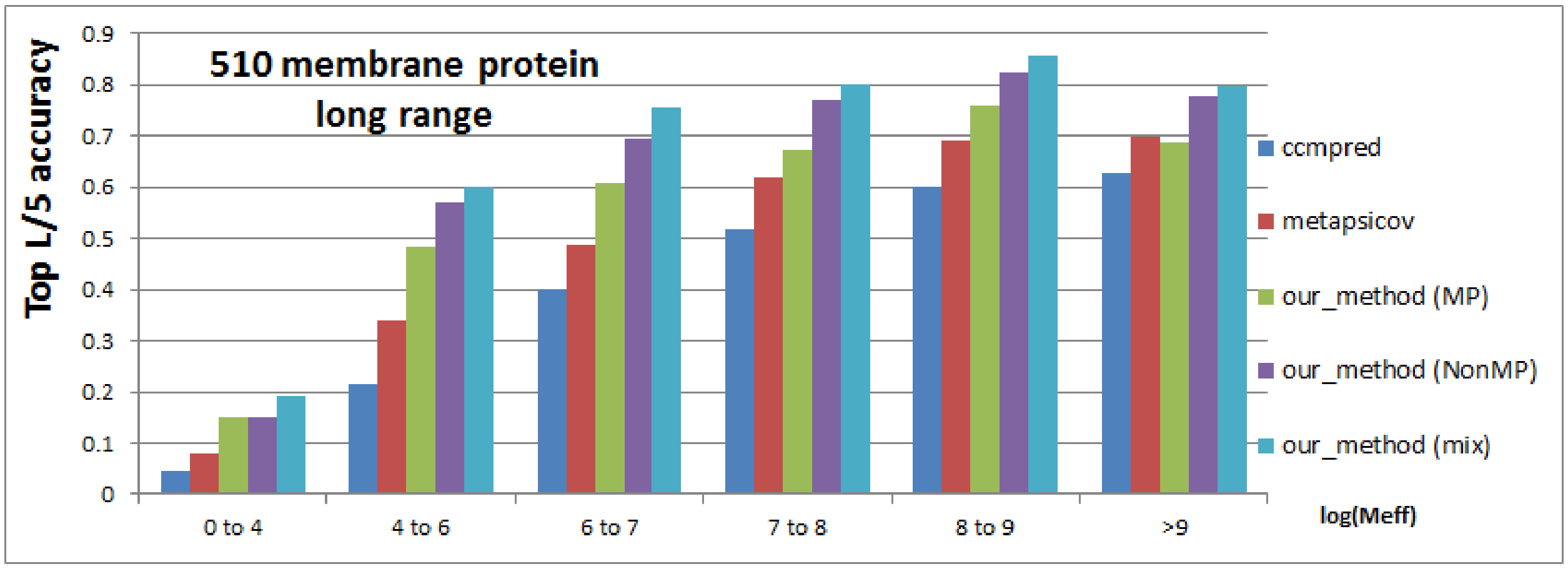

(A)

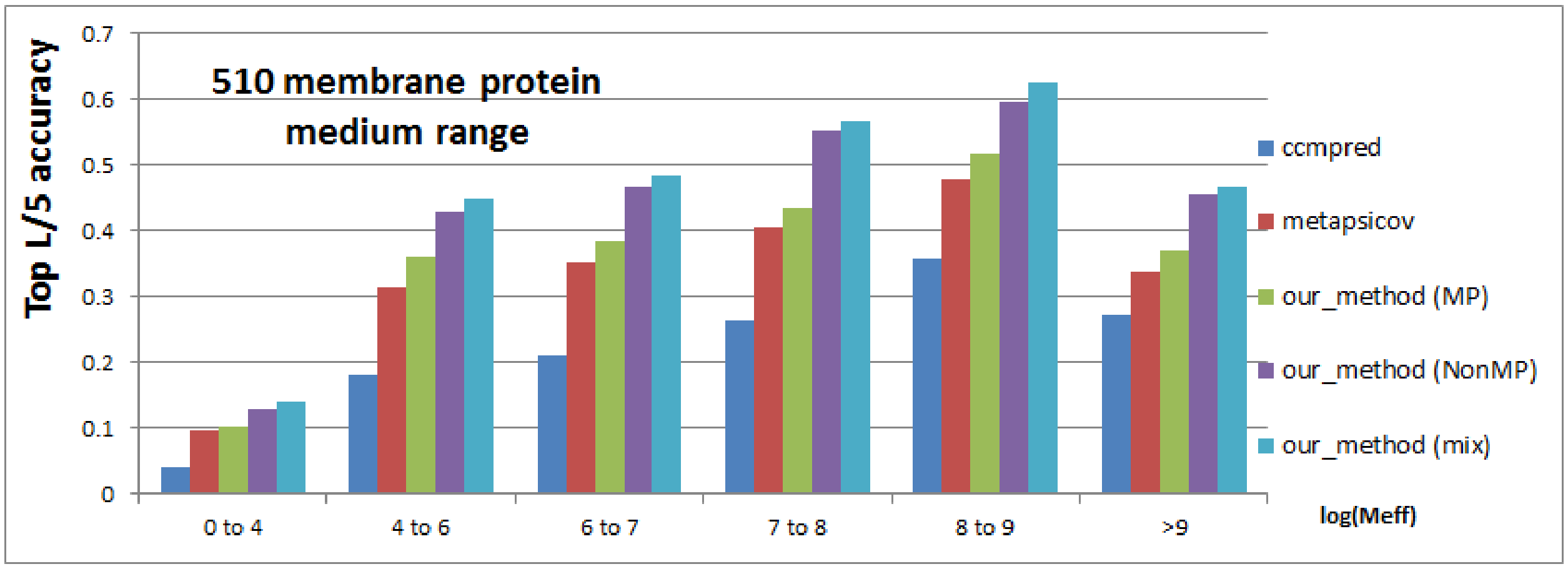

(B)

**Figure 3.** Top L/5 medium-range (A) and long-range (B) accuracy of our three model training strategies Mixed (cyan), NonMP-only (purple), and MP-only (green), CCMpred (blue) and MetaPSICOV (red) with respect to the natural logarithm of the number of non-redundant sequence homologs (X-axis).

## Folding MPs using predicted contacts as restraints

We build 3D structure models for a test MP by feeding its top 2L predicted contacts as distance restraints to the CNS suite [40], which folds proteins from a given set of distance restraints. The average TMscore (RMSD in Å) of the 3D structure models built from our three model training strategies MP-only, NonMP-only and Mixed are 0.45 (14.9), 0.49 (13.2), and 0.52 (10.8), respectively. By contrast, the average TMscore (RMSD in Å) of the 3D structure models built from MetaPSICOV and CCMpred-predicted contacts are 0.39 (16.7) and 0.36 (17.0), respectively. When the best of top 5 models are considered, our three model training strategies can predict correct folds (i.e., TMscore>0.6) for 110, 160, and 200 of 510 MPs, respectively, while MetaPSICOV- and CCMpred models can do so for only 77 and 56 of them, respectively. Figure 4 shows a detailed comparison of our three training strategies, CCMpred and MetaPSICOV when the first 3D models for each test protein are evaluated. See Supplementary Fig. 1 in Appendix for a detailed comparison when the best of top 5 3D models are evaluated. We have also studied the 3D modeling accuracy with respect to the number of sequence homologs available for a protein in question. Fig. 5 shows that our deep models outperform CCMpred and MetaPSICOV no matter how many sequence homologs are available. Our NonMP and Mixed strategies can produce 3D models with TMscore at least 0.1 better than those resulting from MetaPSICOV and CCMpred even if when the protein in question has thousands of sequence homologs.

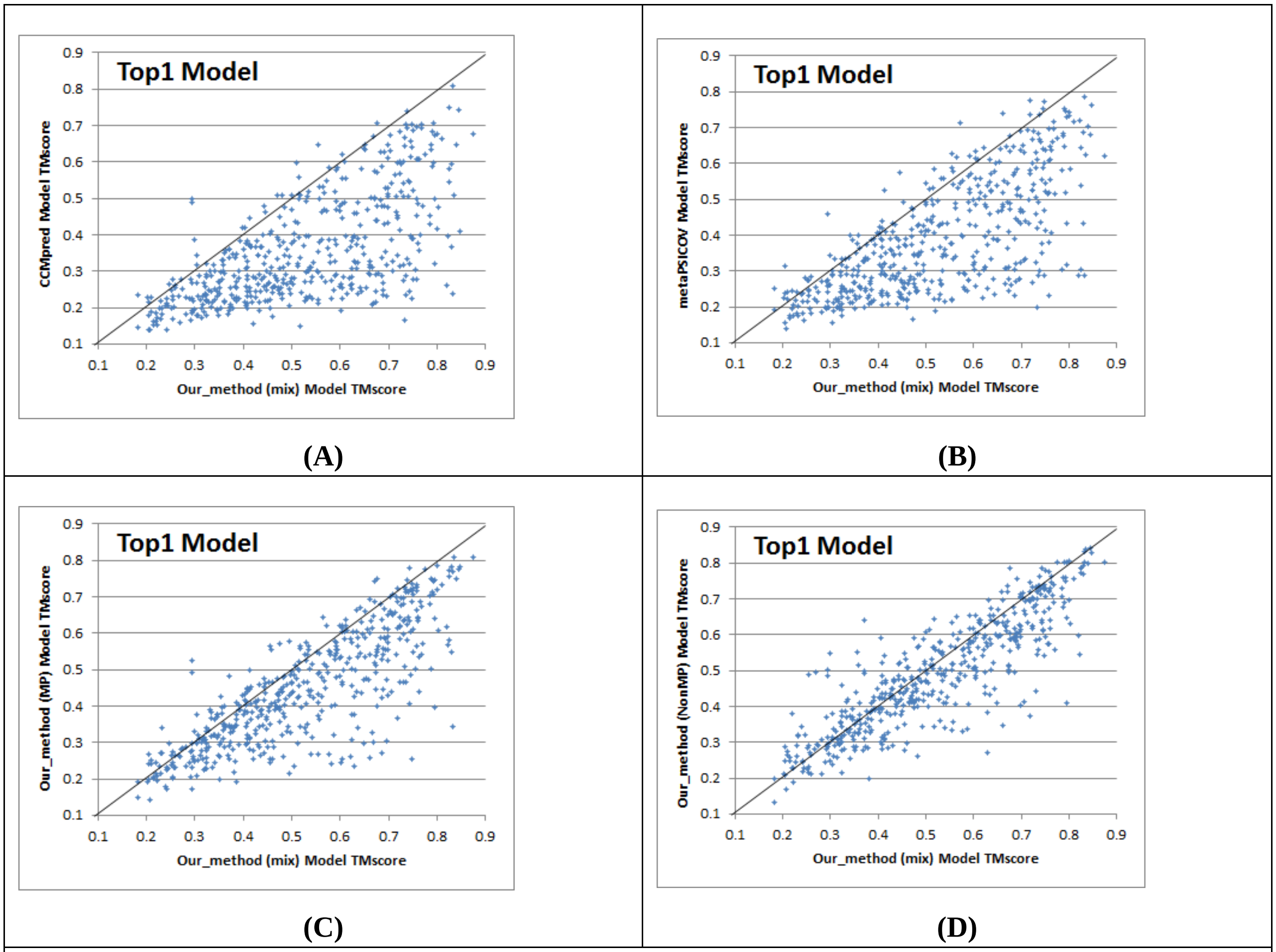

**Figure 4.** Quality comparison (measured by TMscore) of the top 1 contact-assisted models resulting from our methods, CCMpred and MetaPSICOV: **(A)** CCMpred vs. our Mixed strategy, **(B)** MetaPSICOV vs. our Mixed strategy, **(C)** MP-only vs. Mixed, and **(D)** NonMP-only vs. Mixed.

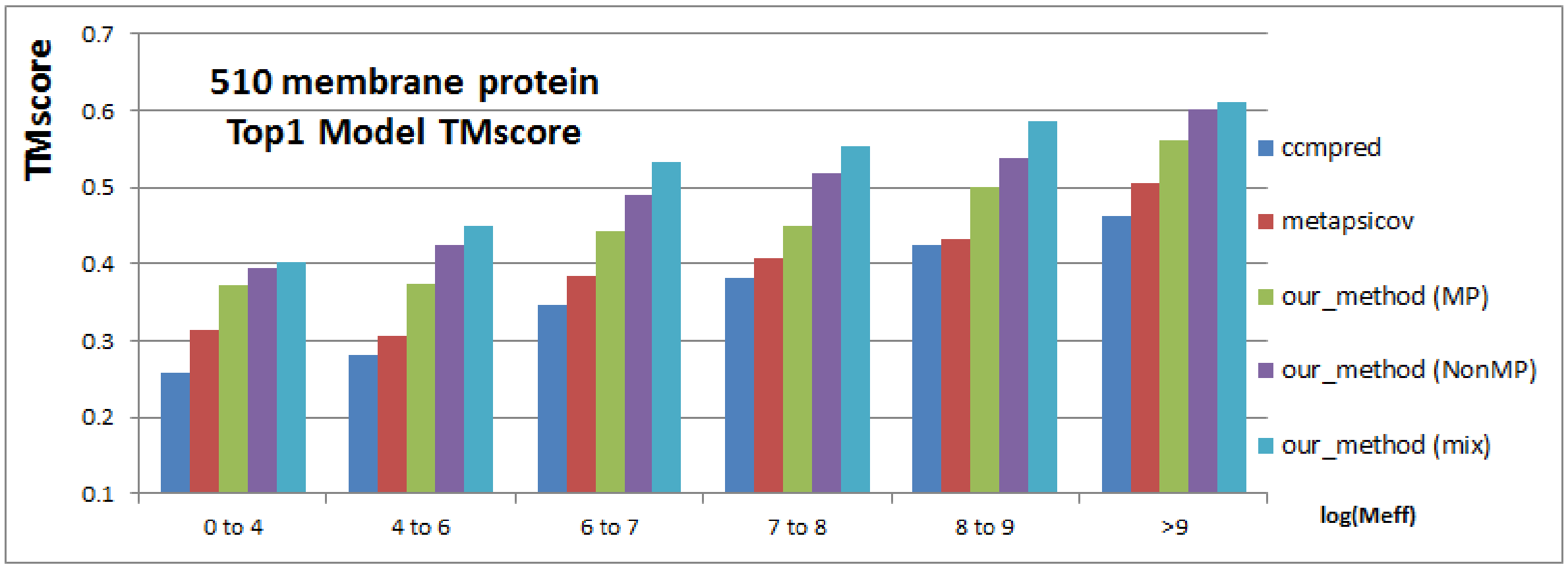

(A)

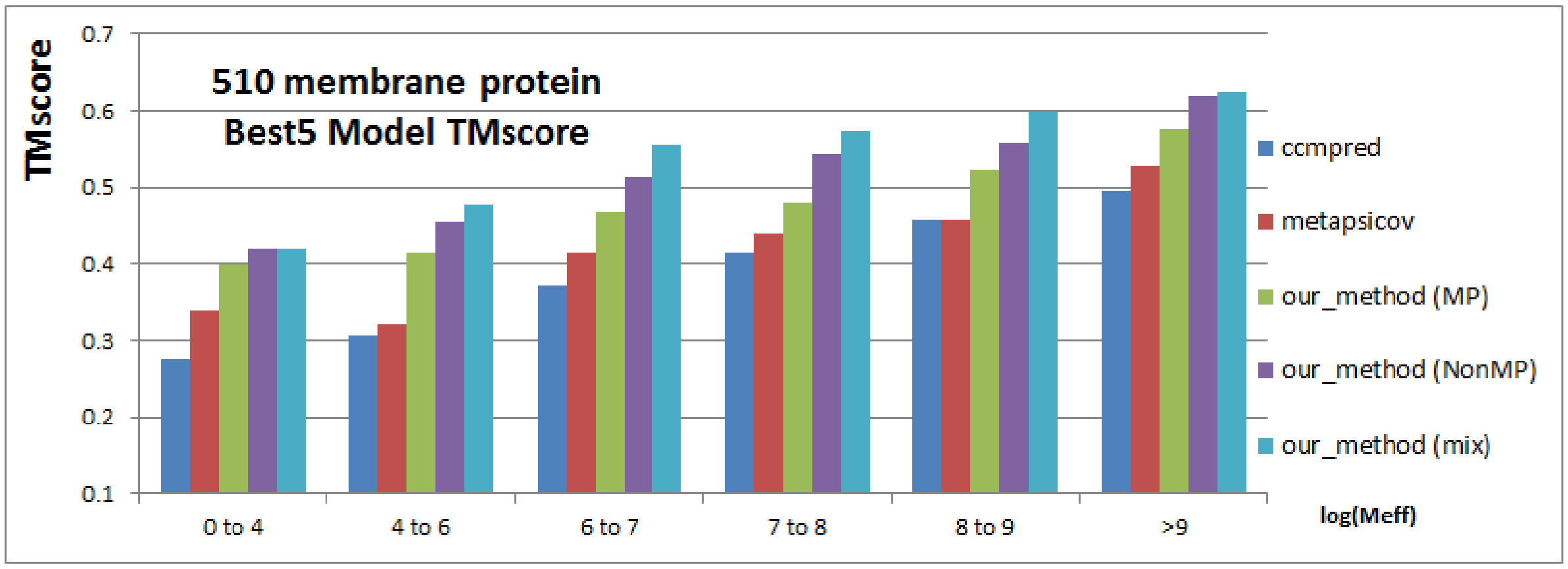

(B)

**Figure 5.** TMscore of the contact-assisted models generated by our three modeling strategies Mixed (cyan), NonMP-only (purple), and MP-only (green), CCMpred (blue) and MetaPSICOV (red) with respect to the natural logarithm of the number of non-redundant sequence homologs (X-axis). The top 1 model **(A)** and the best of top 5 models **(B)** are evaluated on the 510 MPs.

## Comparison with template-based models (TBMs)

To ensure a fair comparison, when comparing TBMs with our NonMP-only strategy, we build TBMs for the 510 MPs only from the 9627 non-MPs (i.e., the training and validation proteins used by our NonMP-only strategy). When comparing TBMs to our Mixed strategy, we build TBMs for the 510 MPs from a mix of the 9627 non-MPs and 4/5 of the MPs (i.e., the training proteins used by our Mixed strategy). For each test MP, we run HHpred [42] to search for the 5 best templates among our training proteins and then construct 5 TBMs by MODELLER [43], which generates 3D structural models by mainly copying from templates. Our Mixed strategy on average has TMscore 0.52 and RMSD 10.8Å and produces 200 3D models with TMscore>0.6. By contrast, the corresponding TBMs have an average TMscore 0.35 and RMSD 17.2Å when only the first models are evaluated. When the best of the top 5 models are considered, the TBMs have an average TMscore 0.38 and RMSD 16.5Å. In total only 40

TBMs have TMscore>0.6, which is much smaller than what can be produced by our contact-assisted folding. Our NonMP-only strategy on average has TMscore 0.49 and RMSD 13.2Å and produces 160 3D models with TMscore>0.6. By contrast, the corresponding TBMs have an average TMscore 0.14 and RMSD 126.3Å when the first models are considered, and the best of the top 5 TBMs have an average TMscore 0.17 and RMSD 111.5 Å. In total only 3 TBMs have TMscore>0.6 since there is very little redundancy between the non-MPs we used and the 510 MPs. All these results suggest that contact-assisted folding may work better for membrane protein modeling when a template with sequence identity >25% is not available in PDB.

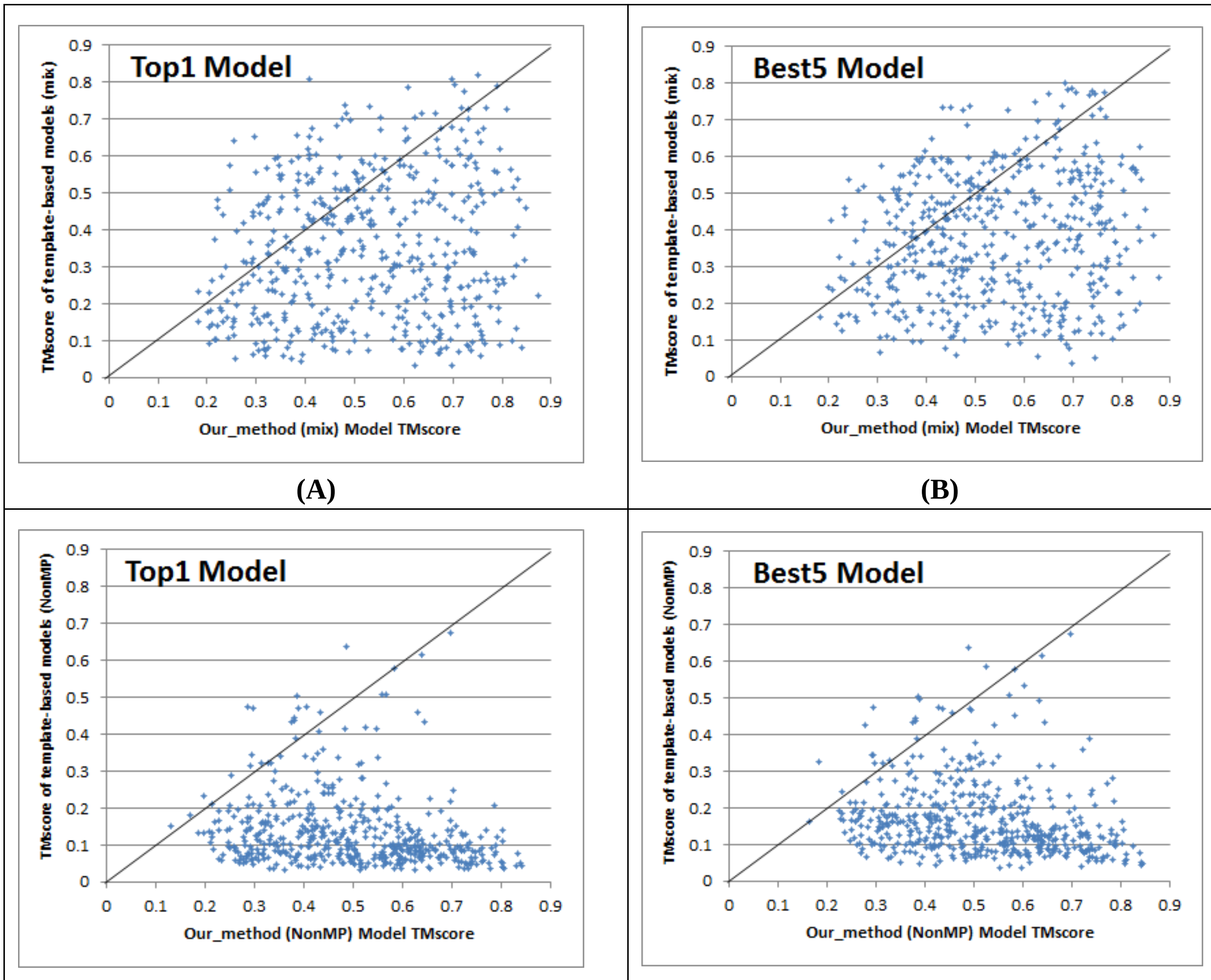

**Figure 6.** Comparison between our contact-assisted 3D models and their template-based 3D models. Two training strategies are compared: Mixed (in upper panel) and from non-MP-only (in lower panel). The top 1 models **(A)** **(C)** and the best of top 5 models **(B)** **(D)** are compared.

## Conclusion

We have presented a novel deep transfer learning method for MP contact prediction and contact-assisted folding that has much better accuracy than existing methods. Our method employs a concatenation of two deep residual neural networks to model contact patterns and sequence-contact relationship. As opposed to existing supervised methods that predict contacts individually, our method predicts all contacts of a protein simultaneously, which makes it easy to model sequence-contact correlation and contact occurring patterns. By making use of a large number of non-MPs in our training set, we can significantly improve the accuracy of MP contact prediction and accordingly MP folding. Our experimental results suggest that even without using any MPs in the training set, our deep learning model still has very good performance, outperforming our deep learning models trained by a limited number of MPs, existing best co-evolution methods and supervised learning methods. This implies that the sequence-structure relationship learned by our model from globular proteins can be well generalized to MP contact prediction.

To understand why our deep model learned from non-MPs works well on MP contact prediction, we examined the contact occurring patterns in both MPs and non-MPs. In particular, we count the occurring frequency of all the $3\times3$ and $4\times4$ contact submatrices and rank them from high to low (see Supplementary Fig. 2 in Appendix). It turns out that the top 20 most frequent contact submatrices of both non-MPs and MPs are quite similar. They differ slightly in the ranking order and accumulative frequency. Even if we consider only MP contact patterns in transmembrane regions, the top 20 most frequent contact submatrices are still the same, but their ranking order has a larger variation. The accumulative frequency of the top 20 $3\times3$ contact submatrix patterns for non-MPs, MPs, and transmembrane regions are 70.5%, 73.4% and 76.3%, respectively. The accumulative frequency of the top 20 $4\times4$ contact submatrix patterns for non-MPs, MPs, and transmembrane regions are 45.6%, 43.9% and 44.4%, respectively. That is, the contact map of non-MPs and MPs is built from a similar set of basic building blocks. This result may partially justify why when non-MPs are used in the training set, we can improve contact accuracy even in transmembrane regions.

We have also studied if MP-specific features such as lipid accessibility [44] and topology structure [45] are useful for MP contact prediction or not. It turns out that by adding them to our deep learning models, we can improve contact prediction accuracy by no more than 1%. This might be because our deep learning models have already implicitly learned these features by themselves. We have also tried other deep transfer learning strategies. For example, we extract the output of the last convolutional layer of two deep models MP-only and NonMP-only, concatenate them into a single feature set and then use this new feature set to train a 2-layer fully connected neural network for MP contact prediction. However, it turns out that this transfer learning strategy cannot outperform our NonMP-only strategy.

Currently we build a 3D model of a test protein using a very simple way. We believe that we shall be able to further improve the 3D modeling accuracy by combining our predicted contacts with fragment assembly and some specific topology information for MPs.

# Appendix

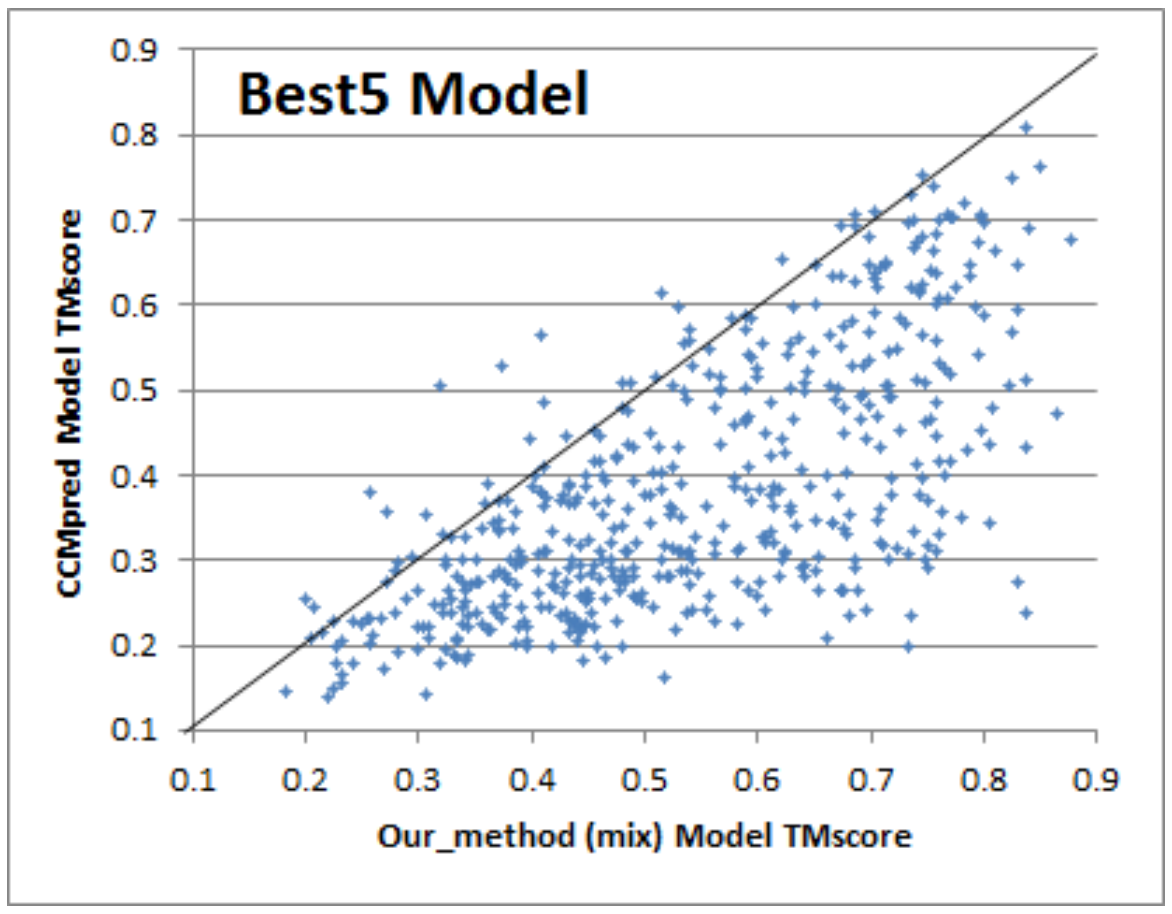

(A)

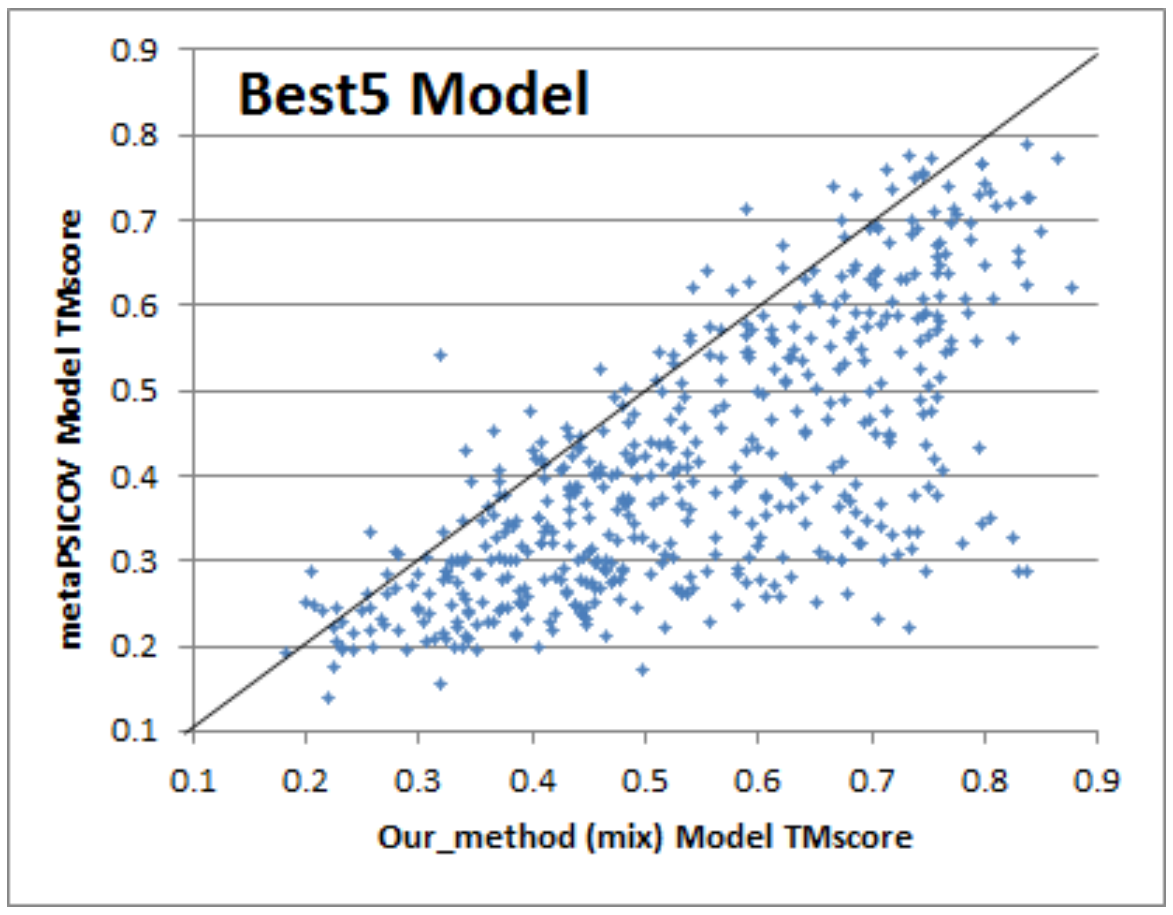

(B)

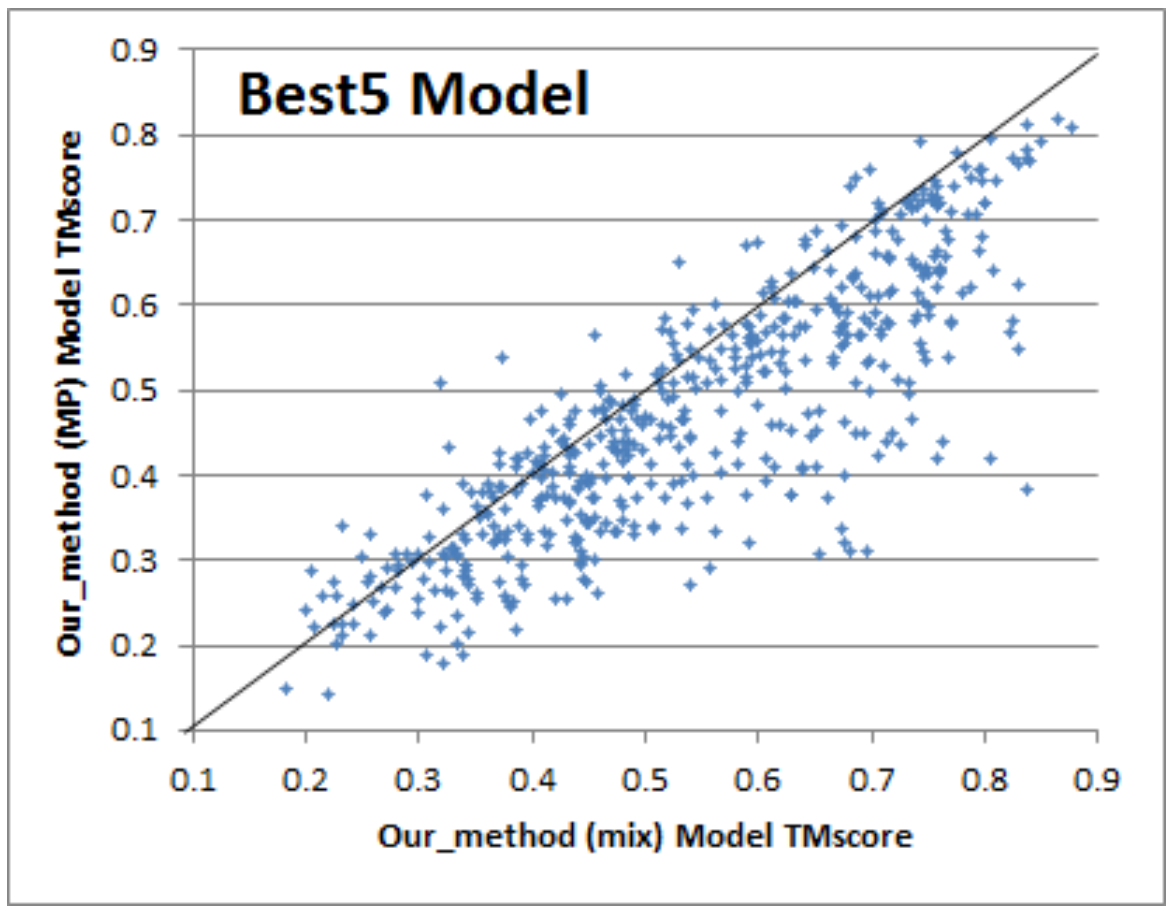

(C)

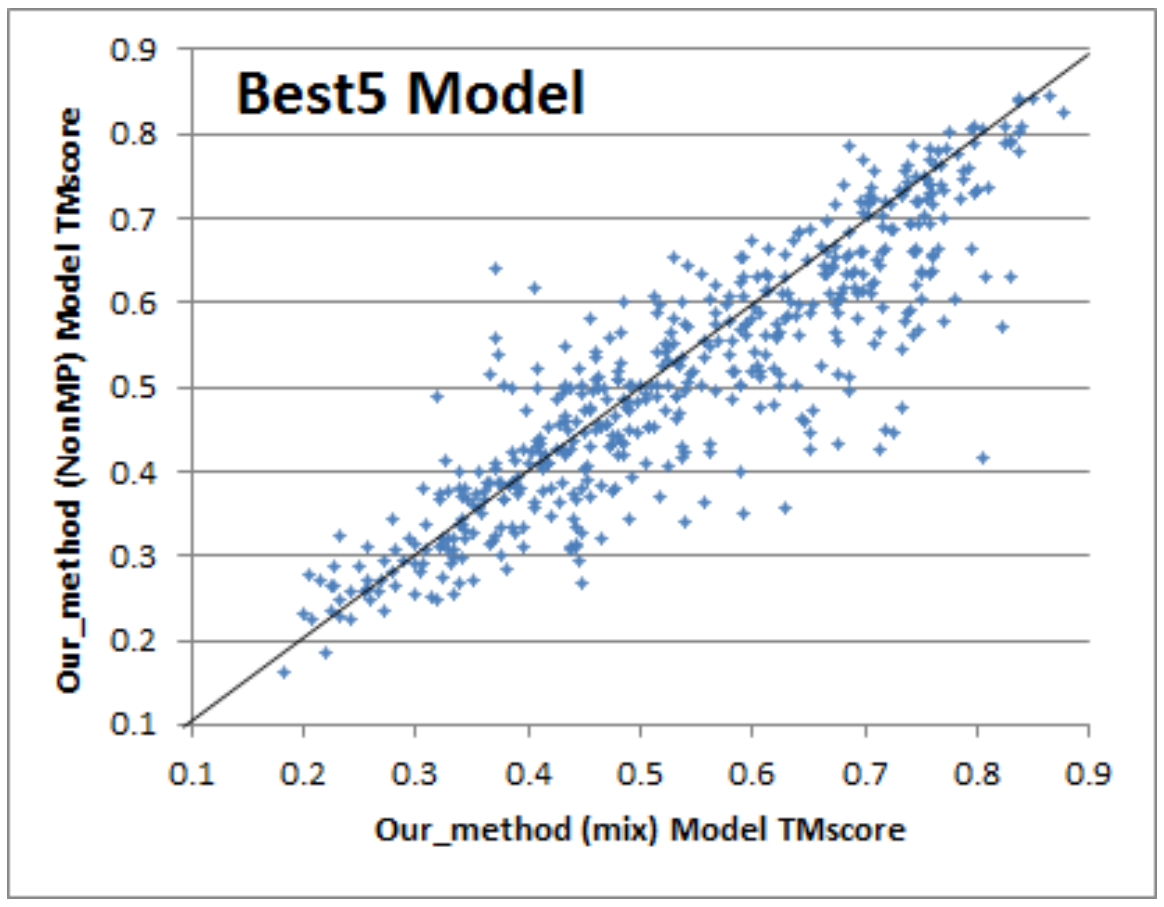

(D)

**Supplementary Figure 1.** Quality comparison (measured by TMscore) of the best of top 5 contact-assisted models generated by our method (trained by a mix of non-MPs and MPs) versus **(A)** CCMpred, **(B)** MetaPSICOV, **(C)** our method trained by MPs only, and **(D)** our method trained by non-MPs only.

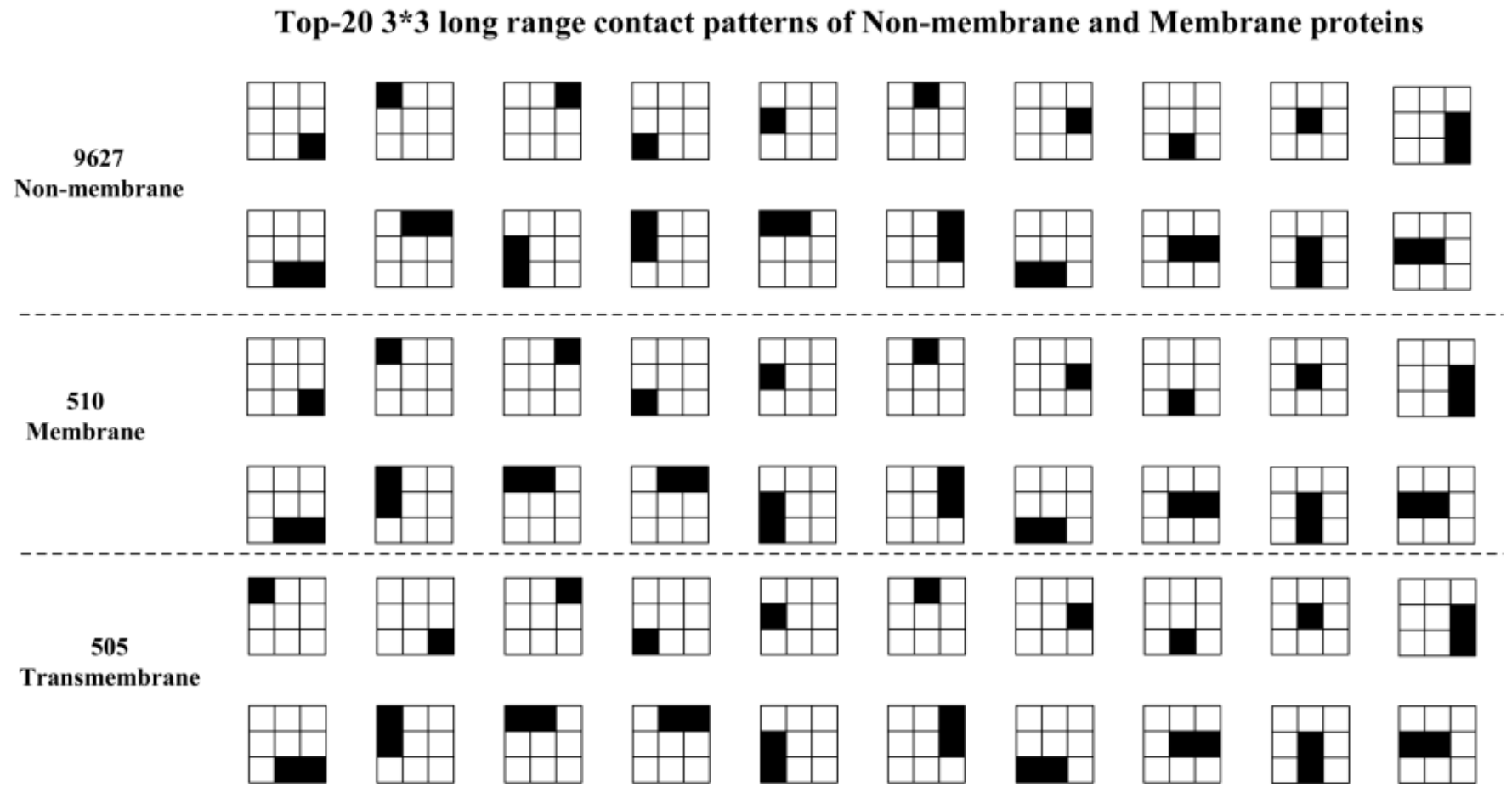

(A)

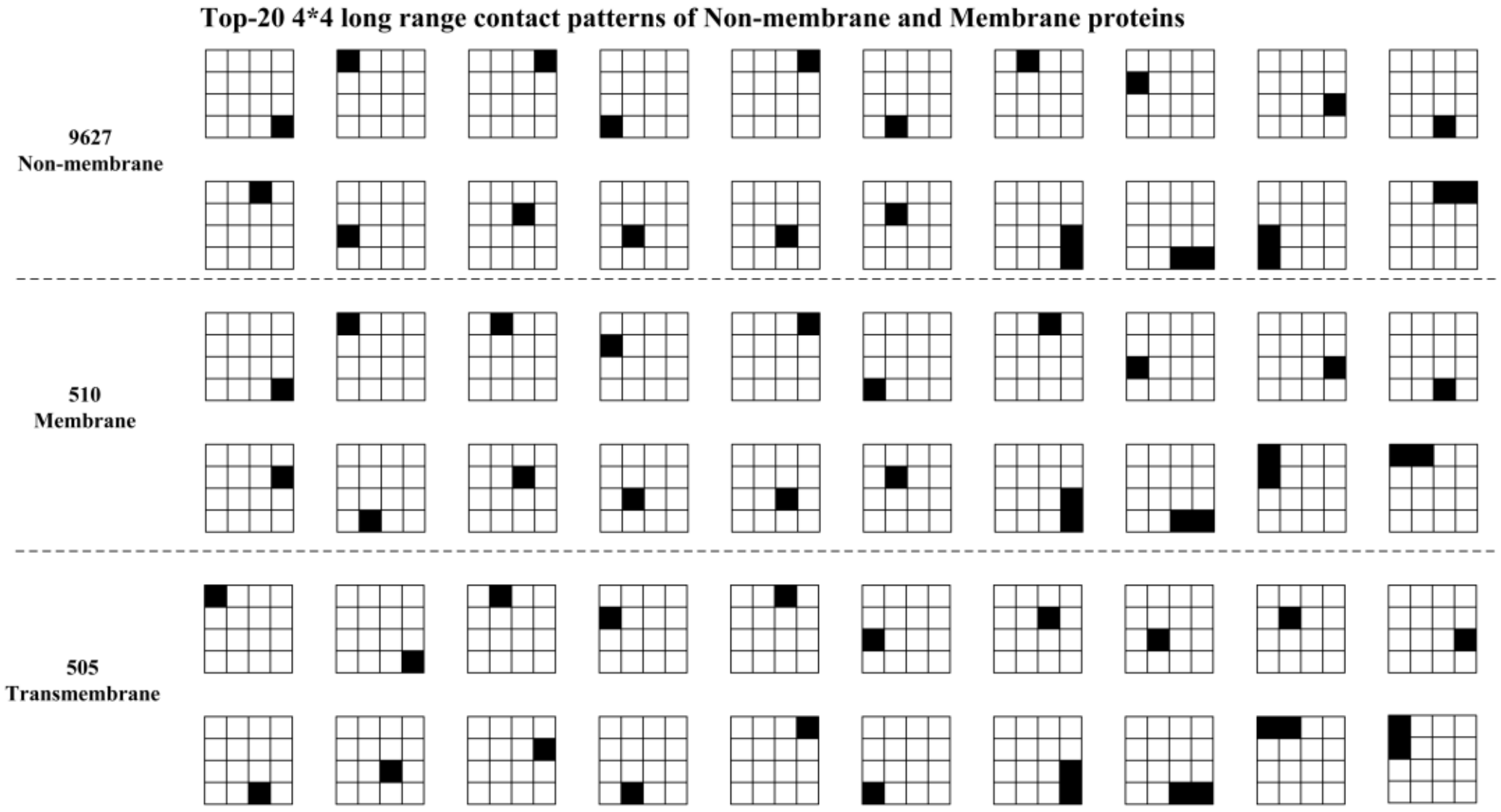

(B)

**Supplementary Figure 2.** Statistics of contact occurring patterns. Both **(A)** 3×3 and **(B)** 4×4 contact submatrices are analyzed. “505 transmembrane” represents only contacts in transmembrane regions are considered.